\documentclass[
]{ceurart}

\usepackage{listings}

\usepackage{longtable}
\usepackage{lscape}
\usepackage[frozencache,cachedir=.]{minted}
\usepackage{cleveref}

\begin{document}

\copyrightyear{2023}
\copyrightclause{Copyright for this paper by its authors.
  Use permitted under Creative Commons License Attribution 4.0
  International (CC BY 4.0).}

\conference{Sem4Tra: Fifth International Workshop On A Semantic Data Space For Transport,
  20 September, 2023, Leipzig, Germany}

\title{Policy Patterns for Usage Control in Data Spaces}

% \tnotetext[1]{You can use this document as the template for preparing your
%   publication. We recommend using the latest version of the ceurart style.}

\author[1]{Tobias Dam}[%
orcid=0000-0002-2463-5831,
email=tobias.dam@fhstp.ac.at
]
\author[2]{Andreas Krimbacher}[%
email=andreas.krimbacher@nexyo.io,
url=https://nexyo.io/,
]
\author[1]{Sebastian Neumaier}[%
orcid=0000-0002-9804-4882,
email=sebastian.neumaier@fhstp.ac.at
]
\address[1]{St.\ Pölten University of Applied Sciences, Austria}
\address[2]{nexyo, Vienna, Austria}

% \author[1,2]{Dmitry S. Kulyabov}[%
% orcid=0000-0002-0877-7063,
% email=kulyabov-ds@rudn.ru,
% url=https://yamadharma.github.io/,
% ]
% \cormark[1]
% \address[1]{Peoples' Friendship University of Russia (RUDN University),
%   6 Miklukho-Maklaya St, Moscow, 117198, Russian Federation}
% \address[2]{Joint Institute for Nuclear Research,
%   6 Joliot-Curie, Dubna, Moscow region, 141980, Russian Federation}

% \author[3]{Ilaria Tiddi}[%
% orcid=0000-0001-7116-9338,
% email=i.tiddi@vu.nl,
% url=https://kmitd.github.io/ilaria/,
% \address[3]{Vrije Universiteit Amsterdam, De Boelelaan 1105, 1081 HV Amsterdam, The Netherlands}

% \author[4]{Manfred Jeusfeld}[%
% orcid=0000-0002-9421-8566,
% email=Manfred.Jeusfeld@acm.org,
% url=http://conceptbase.sourceforge.net/mjf/,
% \address[4]{University of Skövde, Högskolevägen 1, 541 28 Skövde, Sweden}

% %% Footnotes
% \cortext[1]{Corresponding author.}
% \fntext[1]{These authors contributed equally.}

\newcommand{\todo}[1]{\textcolor{red}{#1}}
\newcommand{\sn}[1]{\textcolor{blue}{(sn) #1}}
\newcommand{\td}[1]{\textcolor{teal}{(td) #1}}
\newcommand{\ldk}[1]{\textcolor{orange}{(ldk) #1}}
\begin{abstract}
  Data-driven technologies have the potential to initiate a transportation related revolution in the way we travel, commute and navigate within cities. As a major effort of this transformation relies on Mobility Data Spaces for the exchange of mobility data, the necessity to protect valuable data and formulate conditions for data exchange arises.
  %Our paper presents a collection of policy patterns, which can be used as a foundation for specifying custom policies.
  This paper presents key contributions to the development of automated contract negotiation and data usage policies in the Mobility Data Space. A comprehensive listing of policy patterns for usage control is provided, addressing common requirements and scenarios in data sharing and governance. 
  %Our policy pattern definitions contain the necessary enforcement mechanisms as well as the stakeholder fulfilling the roles of implementation-relevant components of the XACML Policy Enforcement Framework.  
  The use of the Open Digital Rights Language (ODRL) is proposed to formalize the collected policies, along with an extension of the ODRL vocabulary for data space-specific properties.
  %We further demonstrate how to use and combine our policy patterns and provide ODRL examples for each patter via our Github repository.\\
  %\todo{TODO}
\end{abstract}

\begin{keywords}
  usage control policies \sep
  ODRL \sep
  data spaces \sep
  policy enforcement
\end{keywords}

\maketitle

\section{Introduction}
% Mobility Data Space -> Relation?
% related work
% IDS Policy Editor: https://odrl-pap.mydata-control.de/#/
%

The vision of a seamlessly connected mobility ecosystem, where data-driven technologies optimize various aspects of transportation, has the potential to revolutionize the way we travel, commute, and navigate within cities. However, the realization of such a transformative paradigm relies heavily on the establishment of a Data Space for the mobility and transportation domain, wherein data flows effortlessly through fully automated integration processes.
%Nonetheless, the current landscape of multimodal travel information, planning, and booking services is plagued by a pervasive challenge: fragmentation and incompatibility of interchange formats and protocols across different transport sectors. The lack of standardized data formats and interoperability hinder the development of innovative solutions, making it difficult to unlock the full potential of integrated mobility systems. To overcome this hurdle, it is imperative to explore novel approaches that enable efficient data sharing and cooperation between various stakeholders in the transportation domain.

One such effort is the Mobility Data Space (MDS),\footnote{\url{https://mobility-dataspace.eu/}, last accessed 07-07-2023}  founded in 2021 as a non-profit organization by ``DRM Datenraum Mobilität GmbH.'' The MDS serves as a virtual marketplace, facilitating the exchange of mobility data among businesses and government entities. It operates as a decentralized infrastructure, allowing providers to list their data offerings in a catalogue while enabling purchasers to search for specific datasets. Importantly, members of the MDS retain ownership of their data, and the data exchange occurs directly between peers, reducing the involvement of the MDS itself. To facilitate this process, the MDS utilizes the Eclipse Data Space Connector (EDC),\footnote{\url{https://github.com/eclipse-edc/Connector}, last accessed 07-07-2023}  which provides a framework for sovereign, inter-organizational data exchange.

The EDC is a Free and Open Source Software solution developed and released under the Apache 2 License within the Eclipse Foundation. It serves as a concrete implementation of the protocols defined by the International Data Spaces (IDS) standard and aims to ensure compatibility with the requirements of the GAIA-X project. The EDC facilitates the exchange of data through defined data contracts, which are automatically negotiated to regulate access to data assets. 
%The architecture of the EDC emphasizes the separation of the Control Plane and Data Plane. The Control Plane handles tasks related to data management, including data queries, authentication, automatic contract negotiation, and policy enforcement. On the other hand, the Data Plane is responsible for the actual transfer and reception of data, enabling companies to store their data assets in various storage solutions.
While the Mobility Data Space sees the EDC being responsible for automated contract negotiation and the definition and enforcement of usage policies, it is important to note that the EDC is still in an early phase of development at the time of writing. Currently, there is no agreed-upon set of policy patterns implemented by the EDC that supports enforceability by the connector. To contribute to the further development of automated contract negotiation through data usage policies in the Mobility Data Space, we make the following key contributions:
\begin{itemize}
    \item \textit{Collection of Policy Patterns}: Based on a literature review, we identify a number of policy patterns for usage control that are relevant in the context of data spaces, with a specific focus on Mobility Data Spaces. These policy patterns capture common and recurring policy requirements and scenarios encountered in data sharing and governance within the mobility and transportation domain. We discuss concrete examples to illustrate how these identified policy patterns can address real-world requirements. The full list of policies is available online.\footnote{\url{https://github.com/fhstp/dataspaces-policies}}
    \item \textit{Representation using ODRL}: To support implementation and interoperability, we propose the use of the Open Digital Rights Language (ODRL) for representing the collected policies. Since ODRL is a widely adopted language for expressing permissions, obligations, and conditions related to digital rights, we facilitate standardized and machine-readable policy definitions within the context of data spaces. In cases where the ODRL vocabulary cannot be used to express the policy, we propose an extension of the ODRL vocabulary with Data Space-specific properties, available as ODRL Profile.\footnote{\url{https://w3id.org/dataspaces-policies/}}
\end{itemize}
In summary, this paper contributes to the development of data spaces, particularly the Mobility Data Space, by providing a comprehensive list of policy patterns for usage control, exemplifying their practical application, and proposing the use of ODRL for representing these policies.

\section{Data Space Concepts and Policy Framework}

%\subsection{Dataspaces}
% Connectors (EDC): Provider / Consumer

The initial and most significant initiative on data spaces is the International Data Spaces Association (IDSA), initiated in 2015 by the Fraunhofer Society. The non-profit association published a reference architecture \citep{idsram} and an information model \citep{Bader2020TheContent}. 
%The work of the IDSA has been documented in several articles \citep{Otto2019DesigningCase, Zrenner2019UsageEcosystems, Bader2020TheContent, DBLP:conf/isola/PampusJQ22} and an extensive open-access book \citep{2022DesigningSpaces}.
%The IDSA Reference Architecture consists of various components such as brokers, clearing houses, identity providers and app stores. The central component is the connector, which allows to offer and exchange data with other participants. The connectors integrate into each participant's infrastructure, i.\,e.\ the \textit{data source} and \textit{data sink}. Other components of such an IDS ecosystem, which we do not discuss in detail in this paper, include the Clearing House, which serves as a logging service that records information relevant for clearing and billing, and the App Store, which provides apps to the connectors that perform various tasks in the IDS ecosystem (such as transforming, cleansing or analysing data). The tasks of identification, authentication, and authorization are undertaken by an Identity Provider component. 

\begin{figure}
    \centering
    \includegraphics[width=0.8\linewidth]{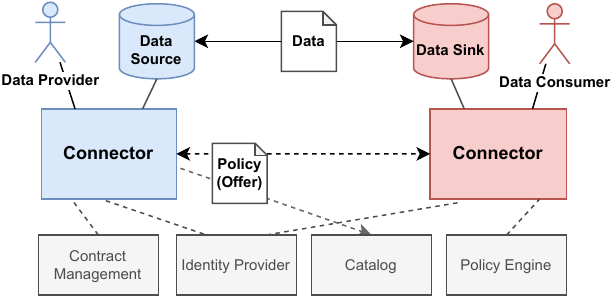}
    \caption{Interactions between the data provider and consumer via connectors and respective components, according to the architecture of the EDC.}
    \label{fig:architecture}
\end{figure}

The Eclipse Dataspace Connector (EDC) project is the most mature instantiation of the IDSA architecture. 
\Cref{fig:architecture} displays the high-level architecture of the EDC components, as described in the connector's documentation.\footnote{\url{https://eclipse-edc.github.io/docs/\#/README}, last accessed 07-07-2023}
In alignment with the IDSA architecture, the connectors require a protocol implementation for usage control policy exchange and enforcement among participants.

\subsection{Stakeholder}
\label{sub:stakeholder}
In a usage control scenario, there are two key entities: the \textit{data provider} and the \textit{data consumer}. In data spaces, these two roles can be interchangeable, as a data consumer may also function as a data provider and vice versa. 

An example in the context of a Mobility Data Space is the scenario where a transportation company ``TransConnect'' collects real-time data from its fleet of vehicles (e.g., GPS coordinates, speed, passenger counts) and shares this data with the analysis firm ``TrafficInsights'', who processes and analyses it (e.g., identifying traffic patterns, congestion hotspots, travel times).
In this scenario, TransConnect acts as the data provider by sharing its real-time vehicle data, while TrafficInsights acts as the data consumer by utilizing the provided data.

However, the roles can be reversed in certain situations: TrafficInsights can act as a data provider by sharing its processed insights and analysis with TransConnect. For instance, TrafficInsights may provide TransConnect with real-time traffic congestion information or recommended alternative routes based on their analysis; TransConnect, in this case, acts as the consumer by utilizing the insights.

In the context of data sharing between a data provider and a data consumer, a trusted \textit{third party} can play a crucial role by providing certificates/guarantees. For instance, consider a data storage provider that acts as a trusted certification authority in our example scenario. Its certificates could verify that TrafficInsights has appropriate data protection protocols in place, such as encryption during data transfer and storage, data anonymization techniques, and adherence to relevant privacy regulations.

\subsection{Enforcement}
\label{sub:enforcement}
Enforcement involves employing mechanisms to ensure compliance with and control the management of usage policies throughout the different phases of usage: before, during, and after. Akaichi and Kirrane \cite{akaichi2022} highlight several important aspects of enforcement:
\begin{itemize}
    \item \textit{Preventive Mechanisms} are dynamic and proactive enforcement mechanisms that allow or prohibit requests for data usage, revoke access in case of policy violations, delay usage requests until obligations are fulfilled, update user or object attributes based on usage decisions, and execute actions like sending notifications to data owners.
    \item \textit{Detective Mechanisms} can be applied in situations where the usage control framework cannot enforce policy restrictions or prevent policy violations. Various detective mechanisms, such as auditing, logging, or user notifications, can be employed to provide evidence or indications of executed commands.
    \item \textit{Continuous Mechanisms} involve managing attributes, conditions, and obligation actions that ensure the validity of ongoing data usage.
    %\item \textit{Conflict Detection and Resolution} addresses the ability of a system to manage conflicting rules in decentralized or distributed systems where data may be subject to multiple policies.
\end{itemize}

\subsection{Components of a Policy Enforcement Framework}
\label{sub:components}

To distribute and modularize the tasks involved in enforcing usage control policies, the framework can be divided into distinct components, where each component focuses on a specific aspect of the process.
The eXtensible Access Control Markup Language (XACML) is an OASIS Open standard designed specifically for Attribute-Based Access Control (ABAC). It provides a declarative fine-grained access control policy language, an architecture, and a processing model for evaluating access requests based on defined policies \cite{oasis2013xacml}. In particular, the core components of the XACML reference architectures are widely used and can also be applied in the context of data spaces to identify the main responsibilities of the stakeholders:

\begin{itemize}
    \item The \textit{Policy Enforcement Point (PEP)} serves as the entry point for enforcement within the XACML framework. It forms a request based on attributes of the requester, the resource being accessed, the requested action, and other relevant information, which is sent to the Policy Decision Point for evaluation. The PEP then enforces the decisions made by the Policy Decision Point \cite{oasis2013xacml}. \emph{In the data space scenario, this point is typically located at the data provider's connector since it administers access to the data source.}
    \item The \textit{Policy Decision Point (PDP)} is responsible for making access control decisions by evaluating the request sent by the PEP and considering the applicable policies. The PDP examines the attributes of the requester, the target resource, and contextual information to determine whether access should be granted or denied. The result of the evaluation is then sent back to the PEP for enforcement \cite{oasis2013xacml}. \emph{Depending on the concrete policy pattern, the data space stakeholder responsible for determining the fulfilment can be either the data consumer or the provider.}
    \item The \textit{Policy Information Point (PIP)} is a component within the XACML framework that provides contextual information during policy evaluation. It supplies the PDP with the necessary context information required to make access control decisions, such as geographical location. The availability of contextual information enhances the granularity of access control decisions \cite{oasis2013xacml}. \emph{In a data space scenario, such contextual information can be either provided by the connector of the provider or consumer, or by a trusted third party.}
    \item The \textit{Policy Administration Point (PAP)} is not directly involved in enforcement but plays a crucial role in the specification and management of usage policies within the XACML framework. The PAP handles the management of policies, including creation, activation, deactivation, and deletion and may provide a user-friendly graphical interface for policy specification and creation \cite{oasis2013xacml}. \emph{In a data spaces scenario, the PAP is located at the stakeholder responsible for deploying the policy.}
\end{itemize}

%PEP: data access is granted via the data provider.
%PDP: based on the policy pattern either at consumer, provider, or third party.
%PIP: information about policy and corresponding descriptive attributes are in catalog of provider and consumer.
%PAP: 

\section{Collection of Policy Patterns}

%In this section we outline the process of collecting applicable policy patterns and provide details on possible implementations. 

%Akaichi and Kirrane~\cite{akaichi2022} provide a survey on different usage control approaches including a list of various usage policy frameworks additionally providing categorization, properties. We We evaluated the original work behind those frameworks and collected policy examples mentioned in the text or if included the concrete instantiation provided in any markup language. We analysed the retrieved policies and aggregated similar ones, filtered those only applicable for a very specific use case or simplified them into general-purpose policies with a wide-ranging applicability. Furthermore, we classified our policies via enforcement type as described in \autoref{sub:enforcement}. For each policy we specified the Policy Decision Point as well as the Policy Administration Point, mentioned in \autoref{sub:components}, and hence specified which stakeholder deploys the policy (PAP) or decides whether it is fulfilled or not (PDP). The stakeholders provider, consumer as well as third party are further described in \autoref{sub:stakeholder}.

%As we compiled our policy collection, we discovered that some use cases presumably relevant for policies used in data spaces were not covered and appended some rules of our own.

In the following we describe the steps applied to collect applicable policy patterns:
\begin{enumerate}
\item \textit{Survey of Usage Control Approaches}: We base our work on the comprehensive survey of different usage control approaches by Akaichi and Kirrane~\cite{akaichi2022}. The survey includes a list of existing usage policy frameworks which we considered for the collection. 
\item Collection of Policy Examples: From these frameworks, we collected any kind of policy examples, patterns and concrete instantiations included.
\item \textit{Policy Analysis and Aggregation}: The collected policies were subjected to analysis and aggregation. Similar policies were identified and grouped together, while policies specific to very particular use cases were filtered out. General-purpose policies with wide-ranging applicability were simplified and included in our final collection.
\item \textit{Classification of Policies}: We classified the policy patterns based on their enforcement type, as described in Section \ref{sub:enforcement}. 
\item \textit{Stakeholder Description}: We identify the stakeholders involved in the implementation of the policy pattern; stakeholders are providers, consumers, and third parties, as described in Section \ref{sub:stakeholder}.
\item \textit{Specification of Policy Information Point and Policy Administration Point/Policy Decision Point}: For each policy pattern in our collection, we specified the corresponding PIP and PAP/PDP, as described in Section \ref{sub:components}. This specification identifies the stakeholder responsible for deploying (PAP) and evaluating (PDP) the policy, and the stakeholder responsible for providing information about its fulfilment (PIP).
\item \textit{Identification of Additional Patterns}: During the compilation of our policy collection, we observed that certain use cases, presumably relevant for policies used in data spaces, were not adequately covered. Consequently, we appended some rules of our own to address these gaps.
\end{enumerate}

% \subsection{Collection}
\autoref{tab:policytable} contains our compiled set of policies, our classifications as well as where they originated from. Our self-defined policies can be identified via the ``-'' in the reference column.

\begin{landscape}
\begin{longtable}[]{@{}p{3cm}p{5.3cm}p{2cm}p{2cm}p{2cm}p{1.2cm}@{}}
\caption{Derived policy patterns classified by enforcement mechanism as well as the respective PIP, PAP/PDP.}
\label{tab:policytable}\\
\toprule
\textbf{Policy Pattern} & \textbf{Description / Example} &
% \textbf{Stakeholder responsibility} 
\textbf{PIP} & \textbf{PAP/PDP}
& \multicolumn{2}{l}{\textbf{Enforcement}}
\tabularnewline
\midrule
\endfirsthead
\endhead
Allow access\strut & Provider allows access to the data for a specific consumer.\strut & Provider\strut & Provider\strut & Preventive\strut & \cite{idspolicies,zhang2008toward}
\tabularnewline
Location / Regional access restriction\strut & Consumer can access data only if located in allowed region.\strut & Provider\strut & Provider\strut & Detective\strut & \cite{idspolicies,zhang2008toward,WuchnerMF13}
\tabularnewline
Location / Regional storage restriction\strut & Consumer can store data only if the storage is located in allowed region.\strut & Consumer, Third Party\strut & Provider\strut & Detective\strut & \cite{idspolicies,zhang2008toward,WuchnerMF13}
\tabularnewline
Time restriction & Consumer can access data only in pre-defined time period.
& Provider\strut & Provider\strut & Preventive & \cite{idspolicies}
\tabularnewline
Access count & Consumer can access the data a fixed number of times. 
& Provider & Provider & Preventive & \cite{idspolicies,zhang2008toward}
\tabularnewline
Rate limit & Consumer can access the data only a limited number of times
within a period. & Provider & Provider & Preventive & \cite{TeigaoMS11}
\tabularnewline
Concurrent active connections & The number of concurrent connections to
retrieve the data is limited. & Provider & Provider & Preventive & \cite{TeigaoMS11}
\tabularnewline
Amount of data & The amount of data that can be transferred/ streamed is
limited. & Provider & Provider & Preventive &  \cite{zhang2008toward}
\tabularnewline
Processing power & The processing power to prepare/ provide the data is
limited. & Provider & Provider & Preventive & \cite{zhang2008toward,CostantinoMMMS18} 
\tabularnewline
Bandwidth & The bandwidth to transfer/ stream the data is limited. 
& Provider, Third Party & Provider & Preventive & \cite{zhang2008toward,CostantinoMMMS18}
\tabularnewline
Billing / Credit points & The consumer will be charged for the data
accesses. & Provider & Provider & Preventive & \cite{idspolicies,zhang2008toward}
\tabularnewline
Data quality & The consumer demands certain data quality standards, e.g., schema conformance, data consistency, etc. & Consumer & Consumer & Detective & -\strut
\tabularnewline
Deletion & Consumer is required to delete data after specific period. & Consumer, Third Party & Provider & Detective & \cite{idspolicies,SchutteB18}
\tabularnewline
Purpose / Application & Consumer is restricted to use data for specific purpose only. & Consumer, Third Party & Provider & Detective & \cite{idspolicies}
\tabularnewline
Provable attribute\strut & Provider demands a certificate/guarantee, e.g. of membership.\strut & Provider, Third Party\strut & Provider & Preventive\strut & \cite{idspolicies,RusselloD09}
\tabularnewline
Encryption by consumer & Provider demands consumer to store data
encrypted. & Consumer & Provider & Detective & \cite{idspolicies,WuchnerMF13}
\tabularnewline
Encryption by provider & Consumer demands provider to transfer data
encrypted. & Consumer & Consumer & Preventive & -\strut
\tabularnewline
Aggregation & Provider demands consumer to aggregate data before
processing. & Consumer & Provider & Detective & \cite{munoz2020data}
\tabularnewline
Anonymization & Provider demands consumer to anonymize data before
processing. & Consumer & Provider & Detective & \cite{munoz2020data,CaoMFC20}
\tabularnewline
Activity logging & Provider demands shared activity logging of the data
processing. & Consumer & Provider & Detective & \cite{idspolicies,SchutteB18}
\tabularnewline
Delegation of permission & Provider demands to attach a certain policy when distributing the data. 
& Consumer & Provider & Detective & \cite{idspolicies,zhang2008toward}
\tabularnewline
Up-to-dateness & Consumer demands that the data is updated with a specified frequency. & Consumer & Provider & Detective & -\strut
\tabularnewline
\bottomrule
\end{longtable}
\end{landscape}

\section{Formalization of Policy Patterns}

This section showcases the applicability of the policy patterns in \cref{tab:policytable} and provides a formalization using the Open Digital Rights Language (ODRL)~\cite{Villata:18:OIM}:

\begin{itemize}
\item The policies follows the ODRL Information Model~\cite{Villata:18:OIM};
\item we make use of the ODRL Core Vocabulary and ODRL Common Vocabulary~\cite{Iannella:18:OVE} to describe the actions and operands of the respective policies, where possible;
\item our representations follow the example use cases from the specification~\cite{Villata:18:OIM};
\item the corresponding Github repository\footnote{\url{https://github.com/fhstp/dataspaces-policies/tree/main/example-policies}} lists a separate ODRL instantiation for each policy pattern;
\item in case we could not model the policy using the ODRL vocabulary terms, we propose extensions in an ODRL Profile. The profile is available in a dedicated ontology.\footnote{\url{https://w3id.org/dataspaces-policies/}}
\end{itemize}

%This section explains the applicability of several policies as defined in \autoref{tab:policytable} and shows how they can be modelled via ODRL. 
%While our Github repository\footnote{\url{https://github.com/fhstp/dataspaces-policies/tree/main/example-policies}} contains a separate ODRL instantiation for each policy pattern, our . 

%All examples are written using Terse RDF Triple Language (Turtle)~\cite{turtle}, follow the ODRL Information Model~\cite{Villata:18:OIM} and utilise the ODRL Core Vocabulary and ODRL Common Vocabulary~\cite{Iannella:18:OVE} as much as possible.

The following examples present selected, combined instantiations of the policy patterns:

% up-to-dateness + data quality\\
% amount of data  + deletion (+ anonymization)\\

\paragraph{Provider-administered policy patterns.}

Listing \ref{lst:limit} combines the ``Amount of data'', ``Deletion'' and ``Anonymization'' patterns of \autoref{tab:policytable}. In this example, the data provider grants the consumer the permission to \textit{read} an asset.\footnote{The W3C's ODRL Vocabulary specification defines the action \textit{read} as ``obtain data from the Asset''~\cite{Iannella:18:OVE}. While this action is a general approach covering any kind of data access, more specific types of access could be defined in a dedicated profile.} In our approach, to be able to limit the amount of data that can be acquired, first the unit of the action \textit{read} is defined to be measured in ``MiB'' (cf. lines 32 to 38). Then the actual limitation is specified via a constraint that requires the count to be less than 1024 (lines 40 to 43). Each execution of the action increases the count, effectively permitting a read of 1 MiB per execution and a total of 1024 MiB. 

Additionally, this policy has two obligations that the consumer has to fulfil. The first obligation (lines 16 to 26) requires the consumer to delete the file before a specific date; the second obligation (26 to 30) requires the consumer has to anonymise the acquired file.

\begin{listing}
    \begin{minted}[frame=single,fontsize=\scriptsize,linenos]{sparql}
@prefix odrl: <http://www.w3.org/ns/odrl/2/> .
@prefix dc11: <http://purl.org/dc/elements/1.1/> .
@prefix xsd: <http://www.w3.org/2001/XMLSchema#> .
@prefix rdf: <http://www.w3.org/1999/02/22-rdf-syntax-ns#> .
@prefix dsp: <http://www.w3id.org/dataspaces-policies/> .

<http://example.com/policies#consumer-administered>
  a odrl:Policy ; odrl:profile odrl:core ;
  odrl:permission _:perm .

_:perm a odrl:Permission ;
  odrl:target <http://example.com/files/file1> ;
  odrl:assigner <https://www.example.com/provider> ;
  odrl:assignee <https://www.example.com/consumer> ;
  odrl:action _:act1 ;
  odrl:constraint _:constr1 ;
  odrl:obligation [
    a odrl:Obligation ;
    odrl:target <http://example.com/files/file1> ;
    odrl:action odrl:delete ;
    odrl:constraint [
      a odrl:Constraint ;
      odrl:leftOperand odrl:dateTime ;
      odrl:operator odrl:lt ;
      odrl:rightOperand "2023-07-10T00:00:00Z"^^xsd:dateTime
    ]
  ], [
    a odrl:Obligation ;
    odrl:target <http://example.com/files/file1> ;
    odrl:action odrl:anonymize
  ] .

_:act1 a odrl:Action ;
  rdf:value odrl:read ;
  odrl:refinement [
    odrl:leftOperand odrl:unitOfCount ;
    odrl:operator odrl:eq ;
    odrl:rightOperand "MiB"
  ] .

_:constr1 a odrl:Constraint ;
  odrl:leftOperand odrl:count ;
  odrl:operator odrl:lteq ;
  odrl:rightOperand "1024" .
  
<http://example.com/files/file1> a odrl:Asset ;
  dc11:title "File 1" .
    \end{minted}
    \caption{The provider limits the amount of data the consumer can acquire, obliges the consumer to delete the data after a specific date as well as requires the consumer to anonymise the data.}
    \label{lst:limit}
\end{listing}

\paragraph{Consumer-administered policy patterns.}

Listing \ref{lst:dataquality} combines the ``Up-to-dateness'' and ``Data quality'' patterns of \autoref{tab:policytable} and gives an example for use cases where the consumer requires the provider to perform specific tasks. I.e., in such a scenario the enforcement of usage control policies is not only relevant for data leaving the provider's domain but also for regulating usage within the consumer's domain. 

\begin{listing}
    \begin{minted}[frame=single,fontsize=\scriptsize,linenos]{sparql}
<http://example.com/policies#consumer-administered>
  a odrl:Policy ;
  odrl:profile <http://www.w3id.org/dataspaces-policies/> ;
  odrl:obligation [
    a odrl:Obligation ;
    odrl:target <http://example.com/files/file1> ;
    odrl:assigner <https://www.example.com/consumer> ;
    odrl:assignee <https://www.example.com/provider> ;
    odrl:action dsp:update ;
    odrl:constraint [
      a odrl:Constraint ;
      odrl:leftOperand odrl:timeInterval ;
      odrl:operator odrl:eq ;
      odrl:rightOperand "P30S"^^xsd:duration
    ]
  ], [
    odrl:action [
      a odrl:Action ;
      rdf:value dsp:qualityControl ;
      odrl:refinement [
        odrl:leftOperand dsp:conformsTo ;
        odrl:operator odrl:eq ;
        odrl:rightOperand <http://example.com/shacl-shape>
      ]
    ] ;
    odrl:constraint [
      odrl:leftOperand odrl:event ;
      odrl:operator odrl:lt ;
      odrl:rightOperand odrl:policyUsage
    ]
  ] .
    \end{minted}
    \caption{The consumer demands the provider to update their data every 30 seconds and requires the data to conform to a specified schema.}
    \label{lst:dataquality}
\end{listing}

Our example (Listing \ref{lst:dataquality}) requires the provider to continuously update the provided data as well as to conform their data to a specified format. Such a conformance specification could be provided using an explicit schema description, e.g., our example uses a Shape Constraint Language (SHACL) schema~\cite{Knublauch:17:SCL}.
The policy is modelled using the terms \textit{update}, \textit{qualityControl}, and \textit{conformsTo} (line 9, 19, and 21) which are self-defined extensions, described in our ODRL profile.\footnote{\url{https://fhstp.github.io/dataspaces-policies/index.html\#update}}

The constraint to the \textit{update} obligation (line 4 to 15) specifies that the action has to be executed every 30 seconds.\footnote{The time interval is specified as a XML Schema Definition (XSD) duration~\cite{Thompson:12:WXS}.}
The \textit{qualityControl} obligation (lines 16 to 25) consists of a refinement to determine what quality control is actually to be applied, i.e., it states that the asset needs to conform to a specific SHACL shape. Additionally, the obligation contains a constraint (line 26 to 30) that requires the asset to conform to the schema before the action is executed.

\section{Conclusion}
The increasing exchange of data in mobility ecosystems, specifically through the use of data space connectors such as the Eclipse Dataspace Connector (EDC), creates the need of automated negotiation of data exchange contracts. To effectively protect valuable data by declaring who can access data under which conditions, participating members of data spaces require policy patterns to implement robust access and usage controls.

To address this need, we have contributed a collection of usage and access control policies, classified according to their enforcement mechanism, and discussed the respective components of the policy enforcement framework in a data space setup. We have demonstrated the application of our policy patterns via example ODRL scenarios for each of the policy pattern. 

In some of the scenarios, we have found that the existing ODRL core and common vocabulary are not sufficient to create ODRL instantiations for the collected policy pattern. To overcome this limitation, we have proposed an initial set of complementing terms as an ODRL Profile;\footnote{\url{https://w3id.org/dataspaces-policies}} this extension needs further definition, extension and evaluation in future work.

While our collection of policy patterns is based on a comprehensive survey of policy frameworks~\cite{akaichi2022}, we acknowledge that it may be incomplete. To provide a more comprehensive set of policy patterns that aligns with the needs of data sharing companies, future research will focus on gathering real-world use cases and obtaining feedback from these companies.

To ensure an efficient adoption of our policy patterns, companies require a complete, well-tested and easy-to-use implementation of a data space connector, e.g. of the EDC. 
Additionally, a graphical user interface to create and administer patterns is essential; there is already existing work in this respect that can serve as a basis.\footnote{IDS Policy Editor, developed by Fraunhofer, \url{https://odrl-pap.mydata-control.de}, last accessed 07-07-2023}

%Additionally, the Eclipse Dataspace Connector needs a framework for validating custom policies according to a defined schema as implemented in the connector.

In conclusion, our research emphasizes the importance of automated negotiation of data exchange contracts and the need for comprehensive policy patterns to enforce usage control in data spaces. By providing a collection of policy patterns and outlining the requirements for their practical application, we aim to contribute to the development of data sharing ecosystems in the context of the Mobility Dataspace.

% conclusions:
% validation of policies according to a defined schema as implemented in the connector

\begin{acknowledgments}
This research was funded by the Austrian Research Promotion Agency (FFG) through BRIDGE project 891103 ``DiDaMe''. The financial support by the Austrian Research Promotion Agency and the Federal Ministry for Digital and Economic Affairs is gratefully acknowledged.
\end{acknowledgments}

%%
%% Define the bibliography file to be used
\bibliography{references}

\end{document}